# Density wave order with antiphase feature associated with the pseudogap in cuprate superconductor $Bi_{2+x}Sr_{2-x}CuO_{6+\delta}$

Zhaohui Wang[1], Han Li[1], Shengtai Fan[1], Jiasen Xu[1], Huan Yang[1,✉] & Hai-Hu Wen[1,✉]

[1]National Laboratory of Solid State Microstructures and Department of Physics, Collaborative Innovation Center of Advanced Microstructures, Nanjing University, Nanjing 210093, China

✉e-mail: huanyang@nju.edu.cn, hhwen@nju.edu.cn

**The strong correlation effect in cuprate superconductors have greatly enriched the phase diagram showing the co-existence of superconductivity with many intertwined orders[1-4]. One of the prominent issues concerning the superconductivity mechanism is about the pseudogap phase which behaves either as cooperator or competitor for superconductivity and its fundamental reason remains still elusive. Here we report the measurements of scanning tunneling microscopy/spectroscopy (STM/STS) in the model superconducting system $Bi_{2+x}Sr_{2-x}CuO_{6+\delta}$ with a transition temperature $T_c$ (zero) $\approx$ 7 K. Although this system is supposed to be slightly overdoped, a pseudogap feature can be easily observed in the energy region of about 20-60 meV. A modulation of local density of states (LDOS) with a periodicity of about $4a_0/3$ ($a_0$: Cu-O-Cu bond length) can be easily observed, which is also supported by the Fourier transformation pattern with wavevectors at about $(0, \pm 3\pi/2a_0)$ and $(\pm 3\pi/2a_0, 0)$. Surprisingly, we find that the LDOS exhibits a clear antiphase feature in the pseudogap energy region below and above the Fermi energy, indicating that it is an intrinsic feature of the pseudogap phase. We interpret this modulation and antiphase feature as a possible consequence of the pair density wave due to the Amperean pairing with finite momentum[5]. Our results give a deep insight on the understanding of the pseudogap phase in cuprate superconductors.**

The microscopic mechanism of high temperature superconductivity remains to be a big puzzle in the field of superconductivity. When hole doping suppresses the antiferromagnetic insulator state in the parent phase of cuprates, superconductivity gradually emerges above a certain doping level, and a pseudogap phase appears at temperatures above the superconducting transition temperature ($T_c$)[6,7]. The superconducting state can also co-exist with charge density wave (CDW) states[8-10]. With more hole doping, the system comes to the overdoped region with a declining of $T_c$, and eventually the "normal state" behaves more or less like a conventional Landau Fermi-liquid behavior in extremely overdoped (OD) region[11]. Numerous studies focused on the underdoped (UD) and optimally-doped (OP) cuprates were conducted and revealed the systematic evolution of the pseudogap and CDWs with the hole doping level $p$. In underdoped region, the nematic electronic structure[12] and $d$-symmetry form factor density wave ($d$FF-DW) [13,14] were observed, both break the rotational (C4) symmetry and translational symmetry simultaneously, and are proved to be coincident with the opening of the pseudogap[15]. In the overdoped region, many interesting and curious phenomena were also observed, such as the $T$-linear resistivity[11,16], the divergence of the normal-state specific heat coefficient[17], the dramatic change of the Hall number from $p$ to $1 + p$ at the pseudogap endpoint[18-20]. Their intimate reasons remain unresolved and are still under hot debate. These peculiar phenomena offer several promising perspectives for understanding the mechanisms underlying the unconventional superconductivity.

To unravel the mystery of mechanism of unconventional superconductivity, a picture concerning pair density waves (PDWs) of two paired electrons was proposed[3-5,21-24]. People used a striped electronic structure within the CuO plane to interpret the 2-dimentional superconductivity in La$_{1-x}$Ba$_x$CuO$_4$ at the magic doping point[24,25], $x$=1/8. Recent experiments have increasingly revealed spectroscopic evidence of PDWs, either as a fundamental state or coexisting with CDWs, across a wide range of correlated materials, including cuprates[26-31], iron-based superconductors[32-34], transition metal dichalcogenides (TMDs) [35,36], kagome materials[37-39], and even heavy-fermion superconductor UTe$_2$[40,41]. The significant correlation between superconductivity and PDWs offers valuable insights into the fundamental mechanisms underlying superconductivity. In cuprates, the PDW state[3,5,21] is widely regarded as a leading candidate for the fundamental order parameter that may characterize the pseudogap. The wavelength of these PDWs, associated with the finite momentum Cooper pairs, extends over several unit cells of the underlying lattice constant, especially $4a_0$ or $8a_0$ in $Bi_2Sr_2CaCu_2O_{8+\delta}$ (Bi-2212) [26-29]. However, it remains uncertain whether the PDW state exists in the overdoped region, and its potential form, if present. If the PDW state does exist, it raises the question of whether this state could account for the enigmatic pseudogap phase.

Here, we focus on crystals of $Bi_{2.08}Sr_{1.92}CuO_{6+\delta}$ (Bi-2201, $T_c \sim 7$ K), and conduct scanning tunneling microscopy / spectroscopy (STM/STS) measurements to visualize the microscopic electronic structure and properties. The La-free Bi-2201 single crystals used here belong to the typical system that shows the super-linear temperature dependence of resistance in wide temperature regime[42]. It is found that the Fourier transformation of the quasi-particle interference (FT-QPI) is suggestive for a closed Fermi surface (FS). Interestingly, one of the main FT-QPI spots associated with the scattering between the diagonal anti-nodal points reveal scattering wave-vectors of $(0, \pm 3\pi/2a_0)$ and $(\pm 3\pi/2a_0, 0)$, which reflects exactly the spatial modulation of local density of states (LDOS) with a period of $4a_0/3$. This modulated electronic state is a feature commonly found in cuprates, ranging from underdoped to overdoped region[43-45]. More remarkable is that, we observed an anti-phased feature of these FT-QPI spots, which is corroborated by the observation of a clear evidence of $\pi$ phase shift for LDOS below and above the Fermi energy in the pseudogap region. Furthermore, it is found that the superconducting gap and related feature also modulate spatially with the same periodicity, which strongly enhances our confidence that the observed modulation is related to the PDW state. We interpret this modulation as the consequence of PDW state induced by the Amperean pairing with finite momentum.

The STM/STS is an ideal technique to detect the microscopic electronic structure, which can allow us to figure out the scattering channels of the quasiparticles. We show the differential conductance map measured at $E = 0$ meV for Bi-2201 in Fig. 1a and its Fourier transform (FT) image in Fig. 1b. The real space QPI image shows clear patches with local $4\sqrt{2}a_0/3 \times 4\sqrt{2}a_0/3$ charge order along nodal direction[46] (highlighted by the purple dashed box) and the ones with local unidirectional bars along antinodal direction with a period of $4a_0/3$ (highlighted by the green dashed box). Thus, the FT-QPI shows eight broad spots at $Q^D \sim (\pm 3\pi/2a_0, 0)$ and $(0, \pm 3\pi/2a_0)$ and $Q^N \sim (\pm 3\pi/4a_0, \pm 3\pi/4a_0)$, which is different from the expectation of the 'octet' model in underdoped or optimally doped cuprates. This observation is in excellent agreement with the FT-QPI pattern of overdoped Bi-2212 ($p \sim 0.23$) [47], which strongly indicates that our sample is indeed locating in the overdoped region. Then we compare our FT-QPI at Fermi energy with the band structure of single layer cuprates[48-50]. Fig. 1c shows the schematic diagram of the band structure of the single layer cuprates with dispersion, and the Fermi surface shape at the bottom. At a low doping, it was shown that hole like Fermi arcs exist with the centers around the Brillouin zone corner ($\pm\pi, \pm\pi$). With increasing hole concentration, the chemical potential drops and eventually passes through the saddle point at ($\pm\pi, 0$) and ($0, \pm\pi$). This results in a Lifshitz transition to an electron-like FS centered at the Brillouin zone center and four intense spots around the antinodal region. Thus, the FS of our Bi-2201 should be electron-like and is schematically plotted in Fig. 1d by the blue outline. We believe unpaired normal quasiparticles exist in present samples, which can give rise to the result of the zero-energy QPI induced mainly by the scattering between the four corners of the FS. It is worth noting that, beside the origin of quasiparticle scattering, the broad spots in the FT-QPI may indicate the short correlation length of the real space modulations. As shown in Supplementary Figure S3, our simulation shows that a shorter modulation correlation length corresponds to larger scattering spots in reciprocal space. Some selective spectra of d$I$/d$V$ vs. bias voltage at various locations are shown in Fig.1e. The corresponding negative second derivation of these spectra are given in Supplementary Figure S1e, revealing the inhomogeneous pseudogap ranging from about 20 meV to 60 meV, but nearly uniform superconducting gap (~12 meV). Then we investigate the property of peak energy, $E_p$, which is calculated by finding the energy of the local maximum of differential conductance on the spectrum at position **r**, here we used the Savitzky-Golay smoothing[51] method to reduce noise in the spectra. The map of $E_p^+(\mathbf{r})$, the peak energy in positive bias region, is presented in Fig.1f, which should consist of both the information of pseudogap and the superconducting gap, as evidenced by the histogram of $E_p$ (Fig.1g). The two-component Gaussian fitting (marked by the red and green solid lines) reveals peak positions near 38 meV, corresponding to the averaged pseudogap, and 13 meV, associated with superconductivity[52-54]. In some regions (~1.5%), we observed the spectrum with the van Hove singularity (vHS) feature, which is consistent with previous reports[55]. The spectral with vHS is also plotted by the yellow curve in Fig.1e. This set of data and analysis reveals intrinsic nanoscale inhomogeneity[54,56,57] of Bi-based cuprate superconductors.

Temperature dependence of tunneling spectra measured at a fixed position for $T$ = 1.5, 3, 5, 7 and 9 K are shown in Fig.1h. Two separated gap features are clearly observed. The coherence peak corresponding to the small gap is clearly suppressed when temperature is raised to about $T_c$ = 7 K, thus it can be ascribed to the superconducting gap; while the large gap is almost temperature-independent indicating the pseudogap feature.

Then, we investigate the phase referenced QPI (PR-QPI), which was first theoretically proposed for a two-gap superconductor[58], and successfully applied to prove the gap-sign-change in iron-based superconductors for non-magnetic impurities[59, 60] and the sign-changing $d$-wave pairing symmetry in optimally doped Bi-2212[61]. The FT-QPI data, which comes from the Fourier transform of the mapping of the differential conductance at a fixed energy $E$, namely $g(\mathbf{q}, E) = |g(\mathbf{q}, E)| \times e^{i\varphi(\mathbf{q}, E)}$, are complex parameters containing the phase information $\varphi(\mathbf{q}, E)$. Then the PR-QPI signal can be extracted from the phase difference between positive and negative bias voltages, which is defined by

$$g_r(\mathbf{q}, E) = |g(\mathbf{q}, +E)| \times e^{i[\varphi(\mathbf{q}, +E) - \varphi(\mathbf{q}, -E)]}. \qquad (1)$$

Fig. 2 presents the QPI data and related PR-QPI analysis. The real space QPI images measured at several specific energies are displayed in Fig. 2a-f, while the corresponding processed PR-QPI images are shown in Fig. 2g-i. In Fig. 2j, we illustrate the energy evolution of the integrated intensity in the areas enclosed by the dashed circles for wavevectors $Q^N$ and $Q^D$ in PR-QPI. At low energies, for example $E$ = 3 meV here, the data reveal no sign change for both $Q^N$ and $Q^D$ (Fig. 2g), which contradicts the expectations for $d$-wave superconductivity[60]. We ascribe this absence of the information of $d$-wave order parameter to the strong impurity scattering against a very small superconducting gap, which yields a very diluted superfluid density. Thus, the QPI consists both the unpaired normal-state quasiparticles and those from very diluted superfluid, and the normal-state quasiparticles play a more important role in the QPI of this sample. In the pseudogap energy region (~40 meV), however, a clear sign-change signal is observed for $Q^D$, while the signal for $Q^N$ becomes very weak without a clear sign-change behavior (Fig. 2h). At higher energies, $Q^D$ signal reverts to a sign-preserved pattern again, as shown in Fig. 2i and Supplementary Figure S4. These PR-QPI signals represent the spatially averaged effect across the whole field of view (FOV), it should pose a corresponding behavior in real space, as we show below.

To gain a comprehensive understanding on the sign-change behavior mentioned above, we perform a more localized and detailed measurement (3×3 nm$^2$) focused on the antiphase modulation along the antinodal direction, the results are shown in Fig. 3. Fig. 3a-f present the differential conductance maps measured at various energies in the same FOV. The corresponding differential conductance profiles along the red arrow in Fig. 3b are respectively shown in Fig. 3g. A visual comparison reveals that the spatial modulations along the red arrow are in phase at $E$ = 100 meV and -100 meV, 10 meV and -10 meV, while the modulations at $E$ = 40 meV and -40 meV are out of phase by π. And these modulations are non-dispersive within a ±100 meV energy window. More clear evidence are presented in Supplementary Figure S5. We term this antiphase modulation of differential conductance in the pseudogap region as APM-PG. The period of APM-PG is $4a_0/3$, which corresponds to the wave vector of $Q^D$. The spectra measured along the red arrow are shown in a color plot in Fig. 3h, which reveals the antiphase feature of the modulations of d$I$/d$V$ at negative and positive bias in the pseudogap region (about 20 to 55 meV). These results demonstrate that a phase difference of π exists between spatial modulations of the filled states and the empty states only at energies with the pseudogap energy scale, whereas the spectra in both higher and lower energy scales beyond the pseudogap region exhibit no such behavior. This shows a striking antiphase feature occurring in the pseudogap regime. In order to know whether this modulation gives influence on superconductivity, we determine the superconducting gaps by taking second derivative of the spectrum (d$I$/d$V$ vs. $V$) and show its spatial dependence in Fig. 3i. Similarly, the relative "coherence peak" determined by $\delta g_\Delta = \delta g_{E=\Delta} - \delta g_{E=0}$ is plotted in Fig. 3j. The raw spectra and the determination of the energy gap as well as $\delta g_\Delta$ are presented in Supplementary Figure S6. Surprisingly, either the superconducting gap or the relative "coherence peak" oscillate in the same period as the APM-PG. The pseudogap peak height, which may reflect the pairing strength, modulates as well (Supplementary Figure S6). And the antiphase pseudogap peak height modulations in filled states and empty states are the key impact for the sign-changing behavior in pseudogap energy region in PR-QPI (Fig.2). Considering the close correlation between the APM-PG, the pseudogap energy and the PDWs[3,21,62], we would like to attribute this APM-PG as the consequence of PDW.

Next, to explore the local unidirectionality of APM-PG, we employ a two-dimensional lock-in technique[13,40] to determine the amplitude and phase of the modulations and then visualize the local unidirectionality. Here we use the differential conductance map at -40 meV (the average value of $\Delta_{PG}$) to visualize the APM-PG, as shown in Fig.4a. We obtain the complex lock-in signal by

$$A_Q^g(\mathbf{r}) = \frac{1}{\sqrt{2\pi}\sigma_r} \int d\mathbf{R} \ g(\mathbf{R}) \ e^{i\mathbf{Q}\cdot\mathbf{R}} \ e^{-\frac{|\mathbf{r}-\mathbf{R}|^2}{2\sigma_r^2}}. \quad (2)$$

so, the amplitude and phase of the modulation are given by

$$\left| A_Q^g(\mathbf{r}) \right| = \sqrt{[\text{Re } A_Q^g(\mathbf{r})]^2 + [\text{Im } A_Q^g(\mathbf{r})]^2}, \quad (3)$$

$$\varphi_Q^g(\mathbf{r}) = \tan^{-1} \frac{\text{Im } A_Q^g(\mathbf{r})}{\text{Re } A_Q^g(\mathbf{r})}. \quad (4)$$

And the local 'magnitude' of APM-PG unidirectionality is then defined as

$$F(\mathbf{r}) = \frac{|A_{Q_{D,x}}^g(\mathbf{r})| - |A_{Q_{D,y}}^g(\mathbf{r})|}{|A_{Q_{D,x}}^g(\mathbf{r})| + |A_{Q_{D,y}}^g(\mathbf{r})|}. \quad (5)$$

Fig. 4b reveals the processed unidirectionality map $F(\mathbf{r})$ for the raw d$I$/d$V$ map measured at -40 meV. When $F(\mathbf{r}) > 0$, represented in red, the APM-PG along x direction dominates; while for $F(\mathbf{r}) < 0$, represented in blue, indicates the dominance of the APM-PG along y direction. Thus, the map of $F(\mathbf{r})$ exhibits spatial heterogeneity, forming a local domain structure of the unidirectional stripy structures. This suggests that the PDW proposed here is microscopically unidirectional, with one direction prevailing within each domain, which is consistent with the PDW state revealed by SJTM[27,30]. The detailed 2D lock-in signal (Supplementary Figure S7) also reveals apparent spatial inhomogeneity and the existence of plenty topological defects. These topological defects may originate from quenched disorders[63] or the strong scattering potential by Bi atoms in the SrO layer due to different radii of Bi and Sr atoms. These perturbations disrupt the formation of long-range order of the density waves (DWs), leading to a short-range feature in the scale of about 3 nm, which is a characteristic of fluctuating DWs[64].

In above, we have shown the clear observation of the short-range APM-PG state with a period of $4a_0/3$. Considering the particular FS of this sample, we propose a logical interpretation for the origin of the observed APM-PG, namely the PDW state. This proposal is inspired by the Amperean pairing scenario[5]. The PDW originates from the pair formed between two fermions inside each high LDOS region at one corner of the electron-like FS, as shown in Fig. 4c. These two fermions carry the momenta of $k$ and $Q^P-k$, respectively. These pairs thus carry a non-zero momentum of $Q^P$, with the preferred periodicity of the PDW being roughly $4a_0/3$. Consequently, this PDW state leads to the exotic density waves with a period of $4a_0/3$ in Bi-2201. We distinguish the $Q^P$ and $Q^D$ because these two wavevectors have different origins, although they actually merge together in this sample. The $Q^P$ doesn't disperse with energy (Fig.3g), which may reflect the intrinsic density wave order associated with the pseudogap; while the $Q^D$ scattering spots may contain contributions both from the band quasiparticle interference and that from the static density wave order, thus the dispersion is also weak (Supplementary Figure S2g). The induced unidirectional APM-PG can be attributed to the coupling between the PDW (with a C4 symmetry) and local inhomogeneity which may have a nematic feature[12,15]. The particle-hole asymmetry of pseudogap predicted by the Amperean pair picture is also observed in our present case, as evidenced by the disparity in pseudogap values and intensities between filled and empty states (Fig. 3h). Additionally, we argue that the APM-PG is related to, but not primarily dependent on the superconductivity, because it occurs in the pseudogap region, rather than in the superconducting gap region. Another strong evidence is that the wave vectors at $(\pm3\pi/2a_0,0)$ and $(0,\pm3\pi/2a_0)$ persist even above $T_c$, according to the previous report[14].

To further illustrate the relevance of this scenario, we review our previous research[65] in underdoped Bi-2201 in which the FS is not closed yet. In the antiferromagnetic region ($p \sim 0.08$), only checkerboard charge order emerges ($4a_0 \times 4a_0$). With more hole doping to the boundary between the antiferromagnetic region and the superconducting dome ($p \sim 0.10$), some unidirectional bars with period of $4a_0/3$ emerge and are confined within the $4a_0 \times 4a_0$ plaquette. It was suggested that two doped holes may accommodate within each plaquette forming the local pairing, but without global phase coherence[65]. In the pseudogap region (~100 meV), a sign reversal feature was also observed, but it is invisible at smaller energies. The results are presented in Supplementary Figure S8. Because the measurement was carried out in a large area with a limited pixel density, thus the data points are not very dense here. But the antiphase feature is quite clear for positive and negative energies. The schematic image of this pairing processes is illustrated in Supplementary Figure S9.

Moreover, other studies revealed that the unidirectional (or bidirectional) $4a_0/3$ modulations were widely witnessed across a wide range of the phase diagram in cuprates[28,43-45,65-67]. While in optimally doped and overdoped Bi-2201, the $4a_0/3$ density waves persist without showing any evidence of the $4a_0 \times 4a_0$ plaquettes[68,69], thus we believe the modulation with a periodicity of $4a_0/3$ observed here is intrinsic, but not a result from the $4a_0 \times 4a_0$ modulations. More strikingly, the compelling Josephson STM experiment on Bi-2212 also reveals the density wave in $|\tilde{D}(\mathbf{q})|$ at $(0, \pm 0.22)2\pi/a_0$ and $(\pm 0.22, 0)2\pi/a_0$ [26], which reflects the spots at $(0, \pm 0.78)2\pi/a_0$ and $(\pm 0.78, 0)2\pi/a_0$ in raw $|\tilde{I}_c(\mathbf{q})|$. Thus, we intend to believe that this novel $4a_0/3$ PDW may also represent the intrinsic form of the pseudogap state.

Next, we discuss whether the reported $4a_0$ $d$FF-DW[13-15] can account for the APM-PG with the wavevectors at $(\pm 3\pi/2a_0, 0)$ and $(0, \pm 3\pi/2a_0)$. Previously, The FT images of the Z-mapping at high energy (above $\Delta_{PG}$) exhibit the spots around $(\pm 3\pi/2a_0, 0)$ and $(0, \pm 3\pi/2a_0)$, which was interpreted as a result coming from the $4a_0$ $d$FF-DW[13-15]. This scenario indeed improves the understanding for the incommensurate modulations, however, the situation is somewhat different in present case observed in Bi-2201. The necessary condition for the $4a_0$ density waves in $k$-space is absent in our overdoped samples (Fig. 1b, d), if we assume these density waves are originated from FS nesting[10]. And we didn't observe any evidence of the $4a_0$ density waves at both high and low energy scales in this sample. We believe the different behaviors between Bi-2201 and Bi-2212 may arise from their distinct band structures[70]. In Bi-2212, the bonding band exhibits much less evolution with hole doping compared to the anti-bonding band. The anti-bonding band undergoes significant changes with hole doping, resembling the behavior shown in Fig. 1c. This band structure of Bi-2212 allows the $4a_0$ density waves to persist across a wide range of hole doping. These analyses indicate that the $4a_0$ $d$FF-DW scenario is not applicable for our present Bi-2201 sample.

Combining our results and the analysis above, we intend to interpret the observed APM-PG effect with the model concerning the PDW state formed by the Amperean pairing, because the wave vector of the two paired electrons within this model is quite close to $(\pm 3\pi/2a_0, 0)$ or $(0, \pm 3\pi/2a_0)$. An alternative picture to explain this APM-PG phenomenon would be the CDW state, formed by the nesting between the momenta with high DOS, such as the corners of the FS (Fig. 1d) with the diagonal scattering between the antinodal regions. But this is unlikely since the APM-PG phenomenon also occurs in very underdoped region in which the Fermi arc terminals are still far away from the antinodal points. Since this PDW state by Amperean pairing induces the gap opening near the antinodal points, it should compete with the Cooper pairing for superconductivity, the latter is formed by the pair-scattering of two electrons with opposite momenta accommodated on the Fermi arcs. We hope our findings provide new insights for understanding this enigmatic behavior in the pseudogap region.

In summary, through careful measurements of scanning tunneling spectra and quasiparticle interference on the self-doped Bi-2201 system, for the first time, we illustrate the sign-changing behavior for the scattering spots $(\pm 3\pi/2a_0, 0)$ and $(0, \pm 3\pi/2a_0)$ in the pseudogap region, which corresponds very well to the modulations of the LDOS with a periodicity of $4a_0/3$ observed in real space. Since this kind of modulation is a ubiquitous phenomenon which exists in cuprate superconductors in wide doping region, we believe it is a common feature that underlies the fundamental physics of the pseudogap region. We interpret this kind of local stripy structure of electronic states to the consequence of pair density wave induced by the Amperean pairing with a finite momentum. The total momentum of the two paired electrons in the Amperean pairing model gives similar wave vector of this PDW order. Our experimental findings undoubtedly reveal incommensurate unidirectional pair density wave modulations along the antinodal direction with an antiphase feature. These findings give a deep insight for understanding the enigmatic pseudogap phase in cuprates, and should help resolve the big puzzle of the pairing mechanism.

# Methods

## Sample synthesis and characterization.

The single crystals of $Bi_{2.08}Sr_{1.92}CuO_{6+\delta}$ were grown by the traveling floating-zone technique under an oxygen pressure P ($O_2$)= 6 atm[71]. The quality of the sample has been checked by the DC magnetization measurement and resistive measurement before the STM measurements. The critical temperature $T_c$ (zero) is about 7 K, and the data were shown in Supplementary Figure S1.

## In-situ STM measurement.

The STM/S measurements were done in a scanning tunneling microscope (USM-1300, Unisoku Co., Ltd.) with ultra-high vacuum, low temperature, and high magnetic field. The Bi-2201 samples were cleaved at room temperature in an ultra-high vacuum with a base pressure of about $2 \times 10^{-10}$ torr and then transferred into the STM system working at 1.5 K. The electrochemically etched tungsten tips were used for all the STM/STS measurements. STM topographies T(r) were acquired in constant current mode, while the tunneling dI/dV spectra and differential conductance maps were determined using the high-precision AC lock-in method, with a small AC modulation voltage at a frequency of 987.3 Hz, and it is approximately proportional to the density of states. All the data were taken at 1.5 K, except for the temperature dependent experiment. The set-point condition of the tunneling junction was declared in the figure legends respectively.

## Lattice Correction in topography $T$(r) and differential conductance maps $g$ (r, E)

The QPI images were measured at series of energies in an appropriate energy window with increments of energy, it takes around 5 days to obtain the dataset. Considering the fact that the spectroscopic imaging STM experiments are time-consuming, the piezo creep (and many other effects) usually causes remarkable distortions on the dataset. Thus we conducted the creep correction on the r-space topography and QPI images to eliminate the creep-caused distortions, assuming that the creep exponentially decays with time[33]. Then the Lawler–Fujita algorithm[12] was used to reduce the local distortions from the non-orthogonality in the x/y axes and obtain more sharp Bragg peaks. This correction can raise the signal-to-noise ratio of FT-QPI patterns and it should not give influence on the final determination of the sign of the PR-QPI signal corresponding to each primary scattering spot.

**Data availability**

Data measured or analyzed for this study are available from the corresponding authors on reasonable request. Source data are provided with this paper.

**Acknowledgements**

We thank H. Luo for the efforts in growing the single crystals, and we also acknowledge helpful discussions with Ilya Eremin, Tetsuo Hanaguri, Ziqiang Wang, Daniel Agterberg, Hong Yao and Zengyi Du. This work was supported by the National Key R&D Program of China (Nos. 2022YFA1403201), the National Natural Science Foundation of China (Nos. 11927809, 12061131001).

**Author contributions**

STM/STS measurements and analysis were performed by Z.W., H.L., S.F., H.Y. and H.H.W. Z.W. and J.X. measured transport properties of samples. H.H.W., H.Y. and Z.W. wrote the paper. H.H.W. coordinated the whole work. All authors have discussed the results and the interpretations.

**Competing interests** The authors declare no competing interests.

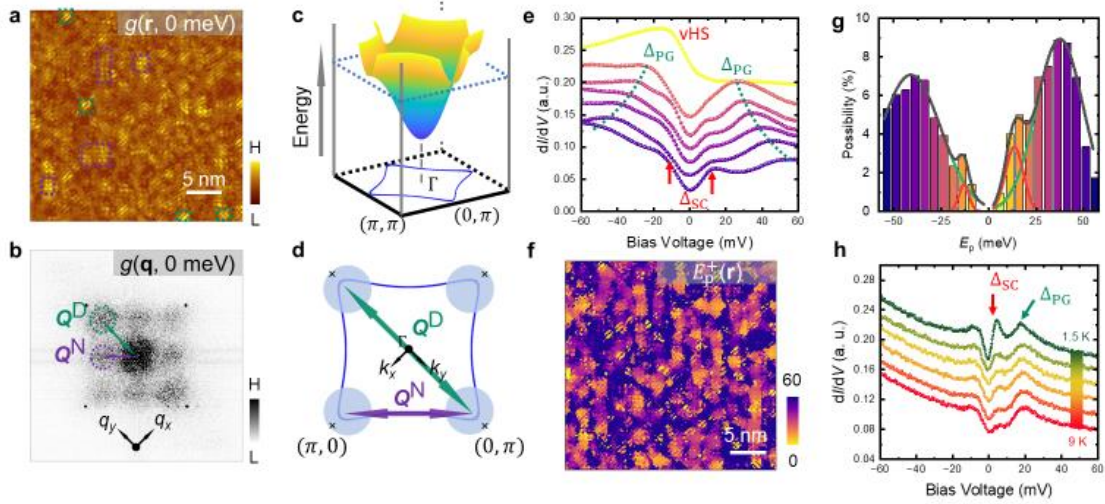

**Figure 1 | Electronic structure and two gap features of Bi-2201. a,** The differential conductance map measured at 0 meV in a 30 nm × 30 nm area. Scale bar, 5 nm. **b,** The FT image of **a**. Four black dots represent the Bragg peaks. Two series of the scattering wave vectors are clearly witnessed, labeled as $Q^N \sim (\pm\frac{3\pi}{4a_0}, \pm\frac{3\pi}{4a_0})$ and $Q^D \sim (\pm\frac{3\pi}{2a_0}, 0), (0, \pm\frac{3\pi}{2a_0})$. **c,** The schematic diagram of the single band structure of single layer cuprate. The parameters for the dispersion shown here are quoted from ref[48]. **d,** Schematic scattering process on the diamond-like Fermi surface obtained from the band structure in **c**. The royal blue shaded areas represent the antinodal region with high density of states. The scattering between these hot spots gives the wave vectors $Q^N$ and $Q^D$. **e,** Examples of d$I$/d$V$ vs. bias voltage spectra taken at various locations in Fig.1a and are offset vertically for clarity. Note that the averaged vHS spectral is also presented, although it is only 1.5% of the area. **f,** The map of peak energy in the positive energy region, $E_p^+(\mathbf{r})$. The peak energy, $E_p$, which is calculated by finding the energy of the local maximum of differential conductance on the spectrum at position **r**, where we used the Savitzky-Golay smoothing[51] method to reduce noise in the spectra. Note that no obvious peak structure is witnessed in the dark purple region. **g,** Histogram of $E_p$ within the same area as **f**. The two-component Gaussian fitting (marked by the red and green dashed lines) reveals peak positions near 38 meV, corresponding to the pseudogap, and 13 meV, associated with superconductivity. The gray solid curves show the sum of these two components. **h,** Temperature dependence of typical tunneling spectral for T = 1.5, 3, 5, 7 and 9 K. Curves are offset vertically for clarity. Two separated gap features are clearly observed. The coherence peak corresponding to the small gap is clearly suppressed when

temperature is raised to about $T_c = 7$ K, while the large gap is almost temperature-independent. The set-point conditions are $V_{set} = -100$ mV, $I_{set} = 100$ pA for **a, e, f**, and $V_{set} = -60$ mV, $I_{set} = 100$ pA for **h**.

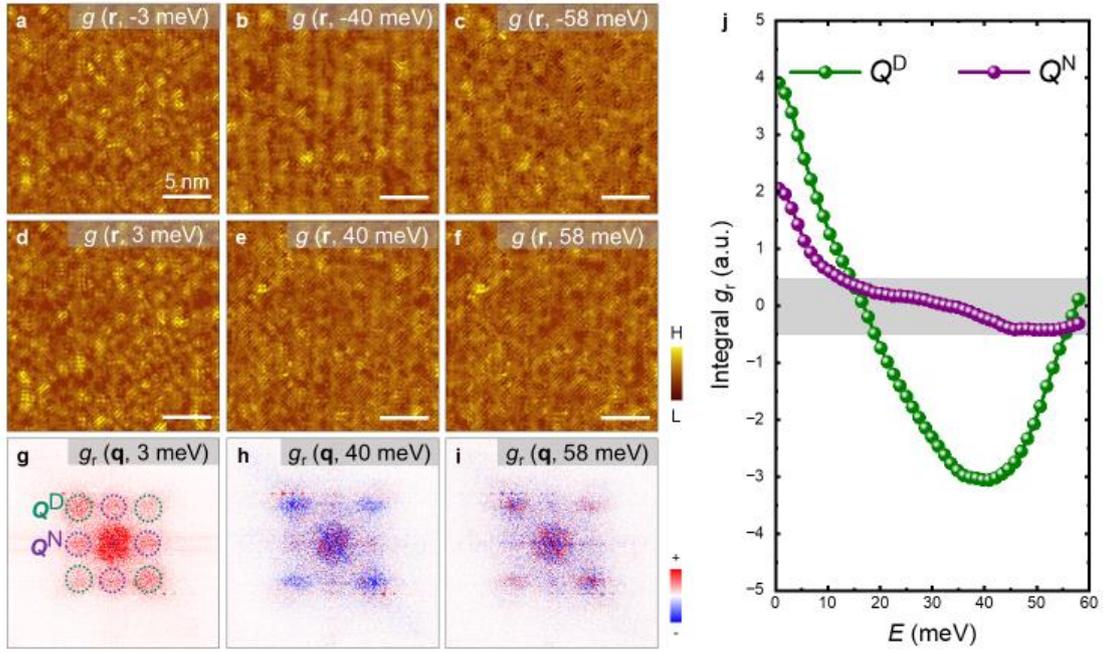

**Figure 2 | Phase referenced QPI analysis. a-f,** Differential conductance mappings at different energies (-3 meV, -40 meV, -58 meV, 3 meV, 40 meV and 58 meV) in the same field of view. Scale bar, 5 nm. **g-i,** Three corresponding phase reference quasiparticle interference (PR-QPI) images obtained directly from the FT-QPI patterns measured at same energies but in filled-states and empty-states. **j,** Energy evolution of integral signal for the two series of wavevectors in the PR-QPI. These peak intensities are obtained by integrating the intensity among the corresponding circle. For the wavevector $Q^D$, clear sign-change behavior is observed in the energy range of 20~55 meV but in higher and lower energy scales beyond the pseudogap region no such behavior exhibits. While for the wavevector $Q^N$, there is no obviously sign-changing signal in the whole energy range within the margin of error. The set-point conditions are $V_{set}$ = -100 mV and $I_{set}$ = 100 pA for measuring all the maps.

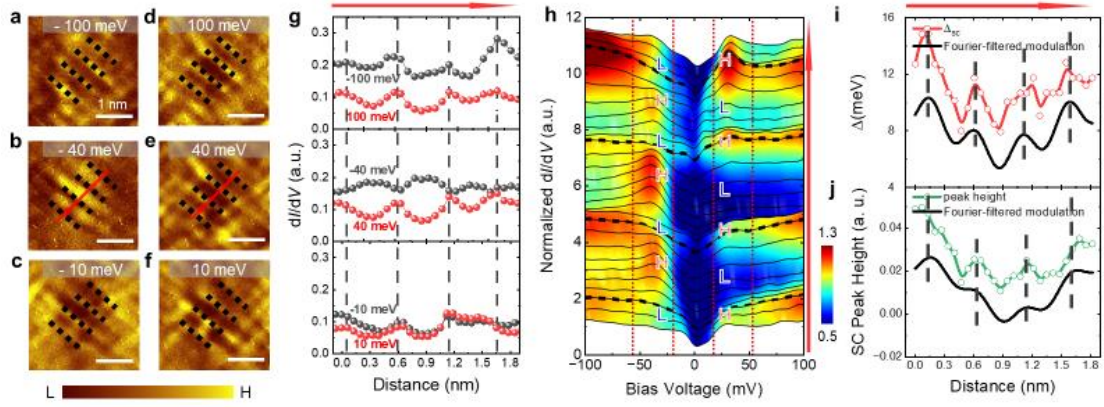

**Figure 3 | detailed measurement for the local PDW state. a-f,** Differential conductance mappings at high energy(±100 meV), pseudogap energy(±40 meV) and low energies (±10 meV) in the same field of view. Scale bar, 1 nm. Black dashed lines are guides to the eye. **g,** Differential conductance profiles along the same line indicated by the red arrow shown in **b** and **e**. The black dashed lines indicate the same location as the guideline in **a-f**. **h,** Color plot of the tunneling spectra measured along the red arrow in **b** and **e**. The black dashed lines indicate the same location as the guideline in **a-f**. The red dashed lines divide the energy region into five parts. In pseudogap energy scale, the antiphase gap feature is clearly observed, as indicated by the antiphase 'H' and 'L'. **i-j,** the superconducting gap and the corresponding peak height modulations extracted from **h,** which show the same period with the APM-PG. The set-point conditions are $V_{set}$ = -60 mV and $I_{set}$ = 100 pA for measuring all the spectra and maps.

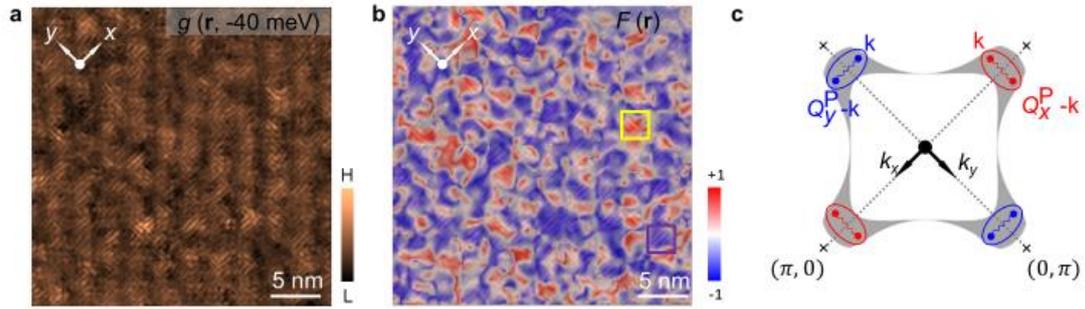

**Figure 4 | Domain analysis for the local APM-PG induced by the PDW state and reasonable pairing process for the PDW. a,** Differential tunneling conductance image $g$ (r, -40 meV) measured below $E_F$ near the pseudogap energy $-\Delta_{PG}$. **b,** Domain configuration of the anisotropy order parameter defined by Eq. (5). This unidirectional PDW regions primarily modulating along y-axis are shaded blue; the regions primarily modulating along x-axis are shaded red. **c,** The schematic diagram of the pairing process in overdoped Bi-2201, based on the model of Amperean pairing.

# Supplementary Information

# Density wave with antiphase feature associated with the pseudogap in cuprate superconductor $Bi_{2+x}Sr_{2-x}CuO_{6+\delta}$

Zhaohui Wang[1], Han Li[1], Shengtai Fan[1], Jiasen Xu[1], Huan Yang[1]✉ & Hai-Hu Wen[1]✉

Fig. S1| Characterization of the self-doped Bi-2201 sample.

Fig. S2| Energy-momentum structure of Bi-2201.

Fig. S3| An alternative interpretation for the large spot in FT-QPI.

Fig. S4| Repeated trials of phase-sensitive experiment with wider energy range.

Fig. S5| Detailed differential conductance profiles along the diagonal direction in Fig.3(a-f).

Fig. S6| Determination for the gap values and corresponding peak height.

Fig. S7| Detailed figures for 2-D lock-in algorithm.

Fig. S8| PR-QPI analysis for our previous data measured on $Bi_2Sr_{2-x}La_xCuO_{6+\delta}$ ($p$~0.1)

Fig. S9| The schematic image of pairing processes for underdoped cuprates.

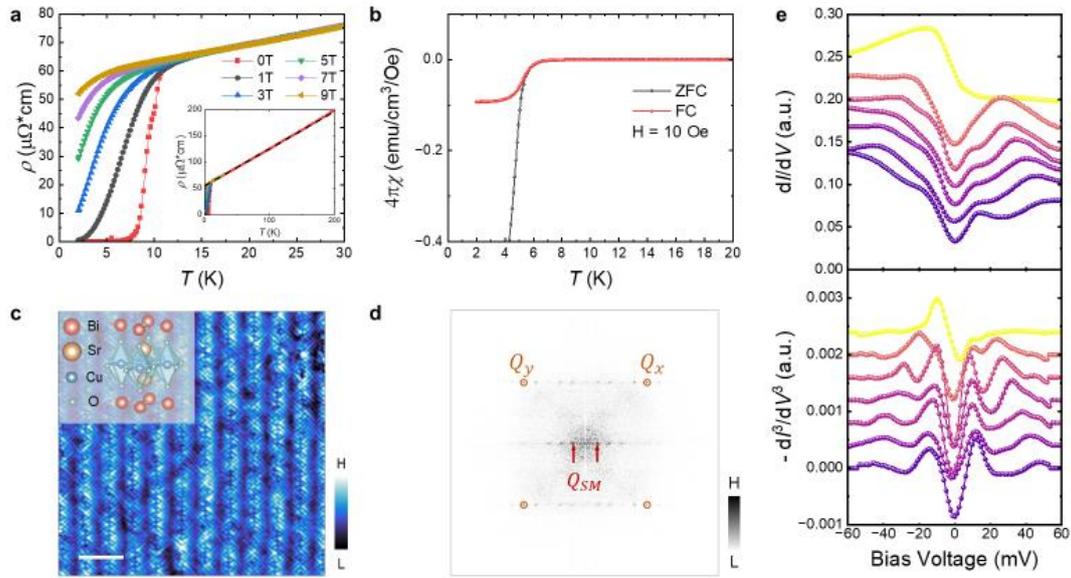

Supplementary Figure S1

**(a)** The resistivity versus temperature curve showing a superconducting transition at 7 K.

**(b)** The temperature dependent susceptibility measured by SQUID.

**(c)** Measured representative topographic image in the same field of view as Fig. 1(a). The creep correction and the Lawler-Fujita algorithm are conducted chronologically to eliminate distortions and enhance the signal-to-noise ratio. The local coupling of electrons to the lattice and the supermodulations in the real-space can be witnessed on the surface. Scale bar: 5 nm. Inset: the schematic diagram of the lattice of Bi-2201, the slightly doped Bi atoms substitute the Sr atoms randomly. The $CuO_2$ plane exists about 6 Å below the BiO plane. Setpoint: $V_{set} = -100$ mV, $I_{set} = 100$ pA.

**(d)** Fourier transform (FT) patterns of (c). The four sharp Bragg peaks, $Q_x$ and $Q_y$, are symbolized by the orange circles. Additionally, the characteristic wave vector of the supermodulations, corresponding to a real-space periodicity of approximately 2.73 nm, is also observed.

**(e)** The same selective STS spectra as shown in Fig.1e and their corresponding negative second derivation. The uniformed superconducting gap and various pseudogap are clearly observed.

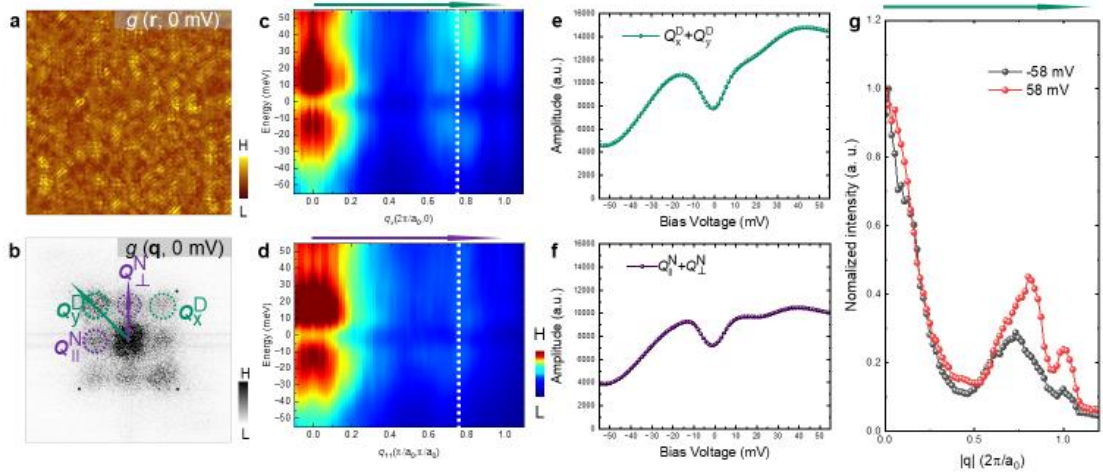

Supplementary Figure S2

**(a,b)** the same figures as Fig. 1(a,b) in main text.

**(c,d)** The energy-momentum structure along the antinodal direction and nodal direction, respectively.

**(e,f)** The evolution of the intensity of two series of the scattering wavevectors with energy.

**(g)** The normalized intensity of line cut along $Q_y^D$ for +58 meV and -58 meV. Locations of the wavevector peaks are mismatched and obey the band dispersion.

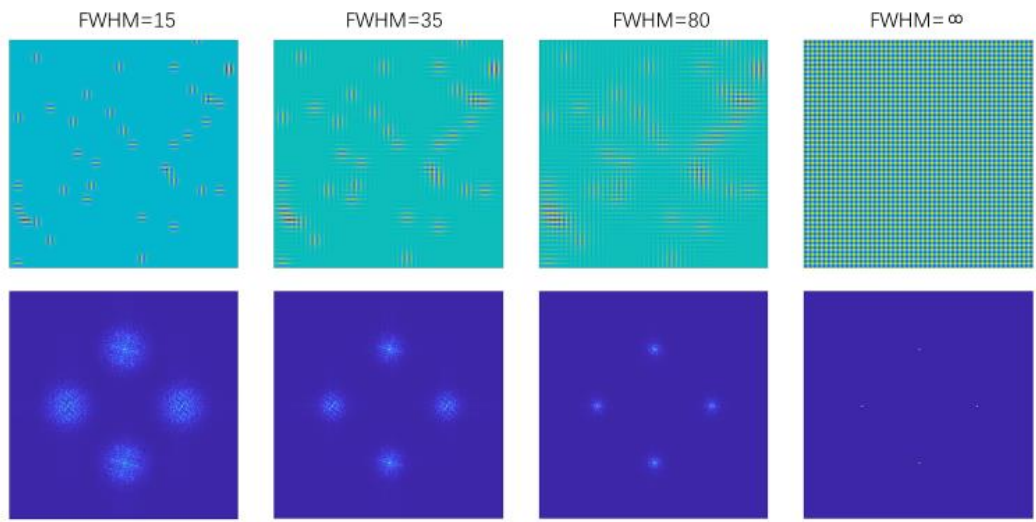

Supplementary Figure S3

Simulated results to interpret the appearance of the board Fourier transforms spots in Fig. 1b. We generate 40 local unidirectional modulations with different full width at half maximum (FWHM) and show their Fourier transform images respectively. Each local unidirectional modulation is the product of unidirectional modulation along $Q_x$ or $Q_y$ and 2-dimentional Gaussian wave packet with the same corresponding FWHM. These results indicate that if the modulation in r-space is short ranged, the Fourier transform spots in q-space will be broad, as the fact that the FWHM of the modulation in r-space and q-space are reciprocal to each other.

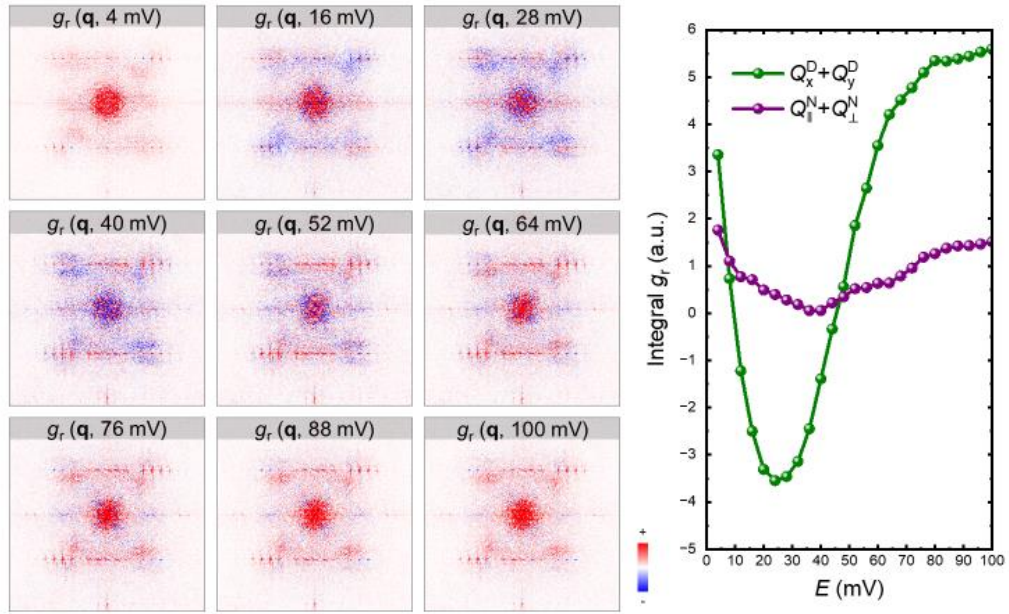

Supplementary Figure S4

repeated trials of phase-sensitive experiment with wider energy range. This experiment was conducted on another sample (with same hole doping) and different STM tip. It shows the same behavior as Fig.2 in the main text. Moreover, the in-phase modulation in high-energy range is observed.

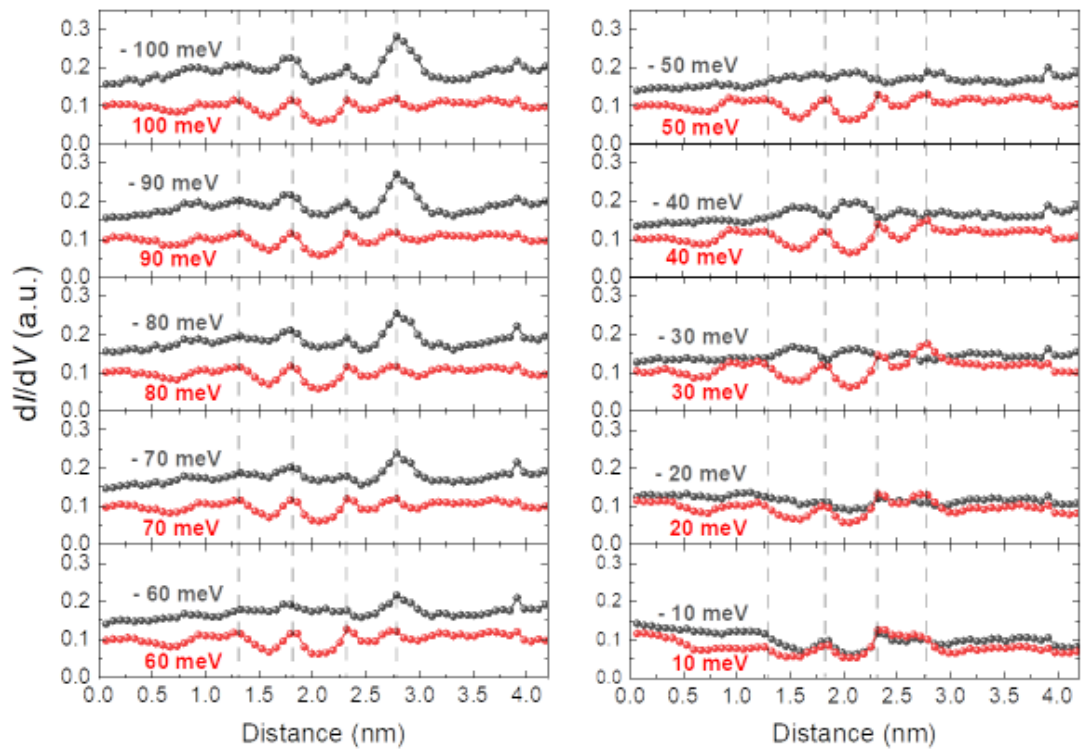

Supplementary Figure S5

Detailed differential conductance profiles along the diagonal direction as the red arrow in Fig. 3(b, e). These results indicate that the out-of-phase modulations arise from variations in modulation amplitude rather than a true phase shift effect.

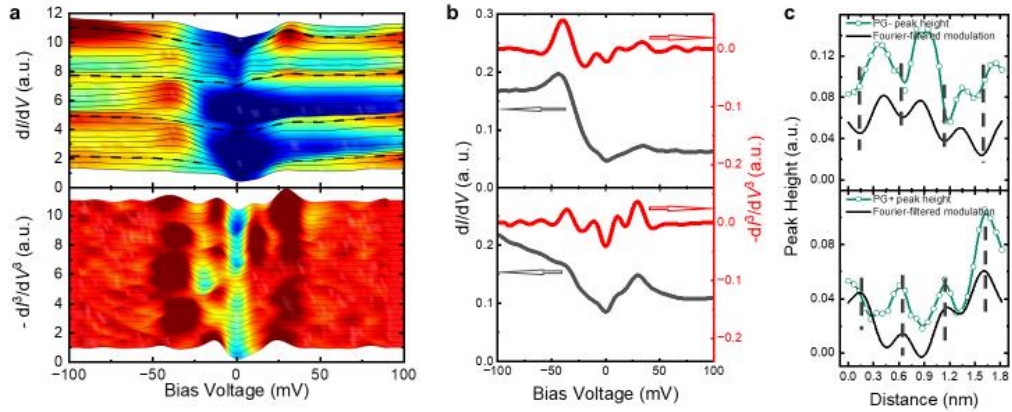

Supplementary Figure S6

(a) The normalized d$I$/d$V$ vs. $V$ spectra and their corresponding negative second derivation. the data are the same as Fig.3h.

(b) Two examples of the d$I$/d$V$ vs. $V$ spectra and their corresponding negative second derivation where we can determinate the gap feature.

(c) the relative peak height determined by $\delta g_\Delta = \delta g_{E=\Delta} - \delta g_{E=0}$ for $\Delta_{PG}$. Antiphase modulations are clear observed.

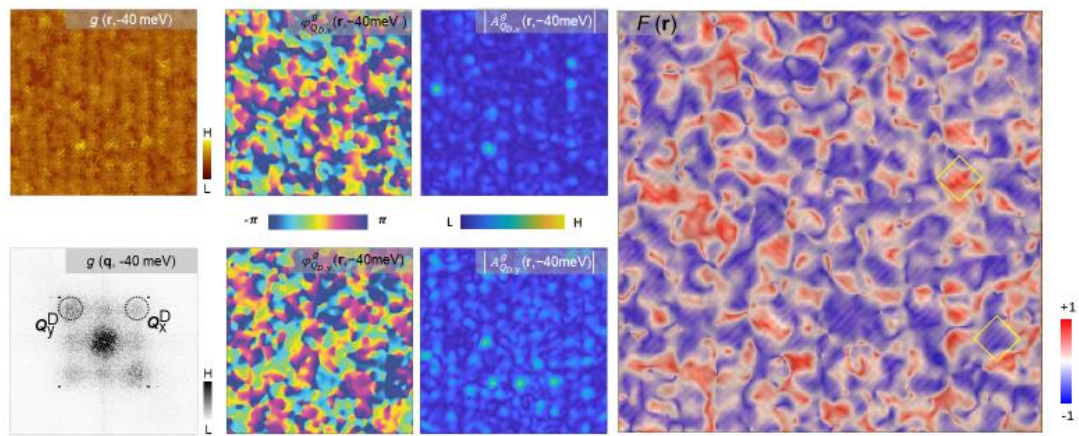

Supplementary Figure S7

Detailed figures for 2-D lock-in algorithm conducted on the differential conductance map $g$ (**r**, -40 meV).

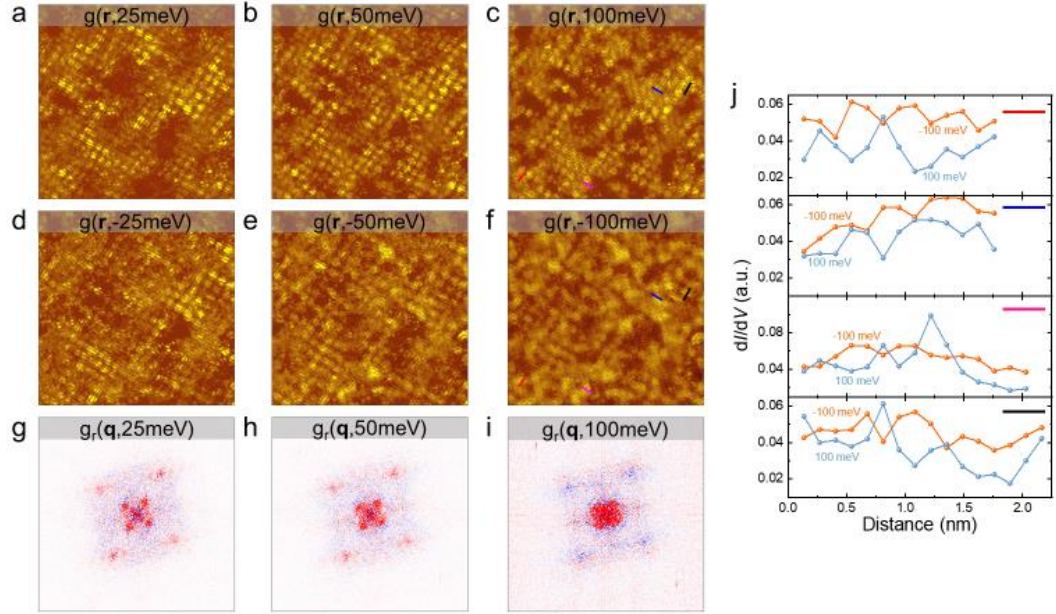

Supplementary Figure S8

(**a-f**) Differential conductance map measured at 25 meV, 50 meV, 100 meV, -25 meV, -50 meV, -100 meV, respectively. These data are measured on the sample $Bi_2Sr_{2-x}La_xCuO_{6+\delta}$ ($p \sim 0.1$), which are quote from ref[65]. The pseudogap was determined as 100-200 meV.

(**g-i**) The PR-QPI for the corresponding energy. In low energy region (20 meV and 50 meV), the PR-QPI images shows sign-preserve signals. While in pseudogap region (100 meV), the sign-changing behavior are clearly observed.

(**j**) Examples for the antiphase modulations at 100meV and -100meV in real space. These line cuts are obtain along the corresponding bars as indicated **c** and **f**

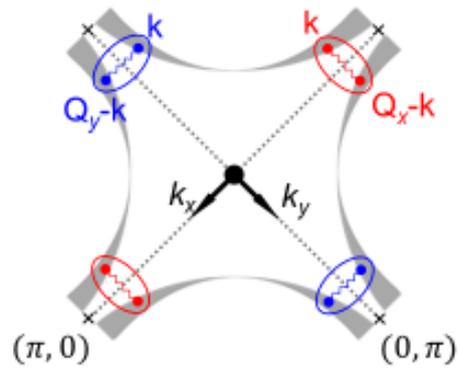

Supplementary Figure S9

The schematic image of the Amperean pairing processes for underdoped cuprates, where the FS is not closed yet.